\documentclass{PoS}

\title{LOFAR: Early imaging results from commissioning for Cygnus A}

\ShortTitle{LOFAR: Early imaging results from commissioning for Cygnus A}

\author{\speaker{John McKean}$^a$, Louise Ker$^b$, Reinout J. van Weeren$^c$, Fabien Batejat$^d$, Laura Birzan$^c$, Annalisa Bonafede$^e$, John Conway$^d$, Francesco De Gasperin$^f$, Chiara Ferrari$^g$, George Heald$^a$, Neal Jackson$^h$, Giulia Macario$^i$, Emanuela Orr\`u$^j$, Roberto Pizzo$^a$, David Rafferty$^c$, Huub Rottgering$^c$, Aleksandar Shulevski$^{c,k}$, Cyril Tasse$^l$, Sebastiaan van der Tol$^{c}$, Ilse van Bemmel$^a$, Ger van Diepen$^a$ and Joris E. van Zwieten$^a$ on behalf of the LOFAR collaboration\\
        $^a$Netherlands Institute for Radio Astronomy (ASTRON), Postbus 2, 7990 AA Dwingeloo, the Netherlands\\
        $^b$SUPA, Institute for Astronomy, Royal Observatory of Edinburgh, Blackford Hill, Edinburgh EH9 3HJ, Scotland\\
        $^c$Leiden Observatory, Leiden University, Postbus 9513, Leiden, 2300 RA, the Netherlands\\
        $^d$Onsala Space Observatory, SE-43992 Onsala, Sweden\\
        $^e$Jacobs University of Bremen, Campus Ring 1, 28759 Bremen, Germany\\
        $^f$Max-Planck-Institute for Astrophysics, Karl Schwarzschild Str. 1, 85741 Garching, Germany\\
        $^g$UNS, CNRS UMR 6202 Cassiop\'ee, Observatoire de la C\^ote d'Azur, Nice, France\\
        $^h$Jodrell Bank Centre for Astrophysics, School of Physics and Astronomy, The University of Manchester, Oxford Road, Manchester M13 9PL\\
        $^i$INAF - Istituto di Radioastronomia, via Gobetti 101, 40129 Bologna, Italy\\
        $^j$Radboud University Nijmegen, Heijendaalseweg 135, 6525 AJ Nijmegen, the Netherlands\\
        $^k$Kapteyn Astronomical Institute, University of Groningen, P.O.Box 800, 9700 AV Groningen\\
        $^l$GEPI, Observatoire de Paris-Meudon, 5 place Jules Janssen, 92190 Meudon, France\\
        \\
        E-mail: \email{mckean@astron.nl}}

\abstract{The Low Frequency Array (LOFAR) will operate between 10 and 250 MHz, and will observe the low frequency Universe to an unprecedented sensitivity and angular resolution.  The construction and commissioning of LOFAR is well underway, with over 27 of the Dutch stations and five International stations routinely performing both single-station and interferometric observations over the frequency range that LOFAR is anticipated to operate at. Here, we summarize the capabilities of LOFAR and report on some of the early commissioning imaging of Cygnus A.}

\FullConference{10th European VLBI Network Symposium and EVN Users Meeting: VLBI and the new generation of radio arrays\\
		September 20-24, 2010\\
		Manchester UK}

\begin{document}

\section{The Low Frequency Array (LOFAR)}

\begin{figure}
\begin{center}
\includegraphics[angle=0,width=13.5cm]{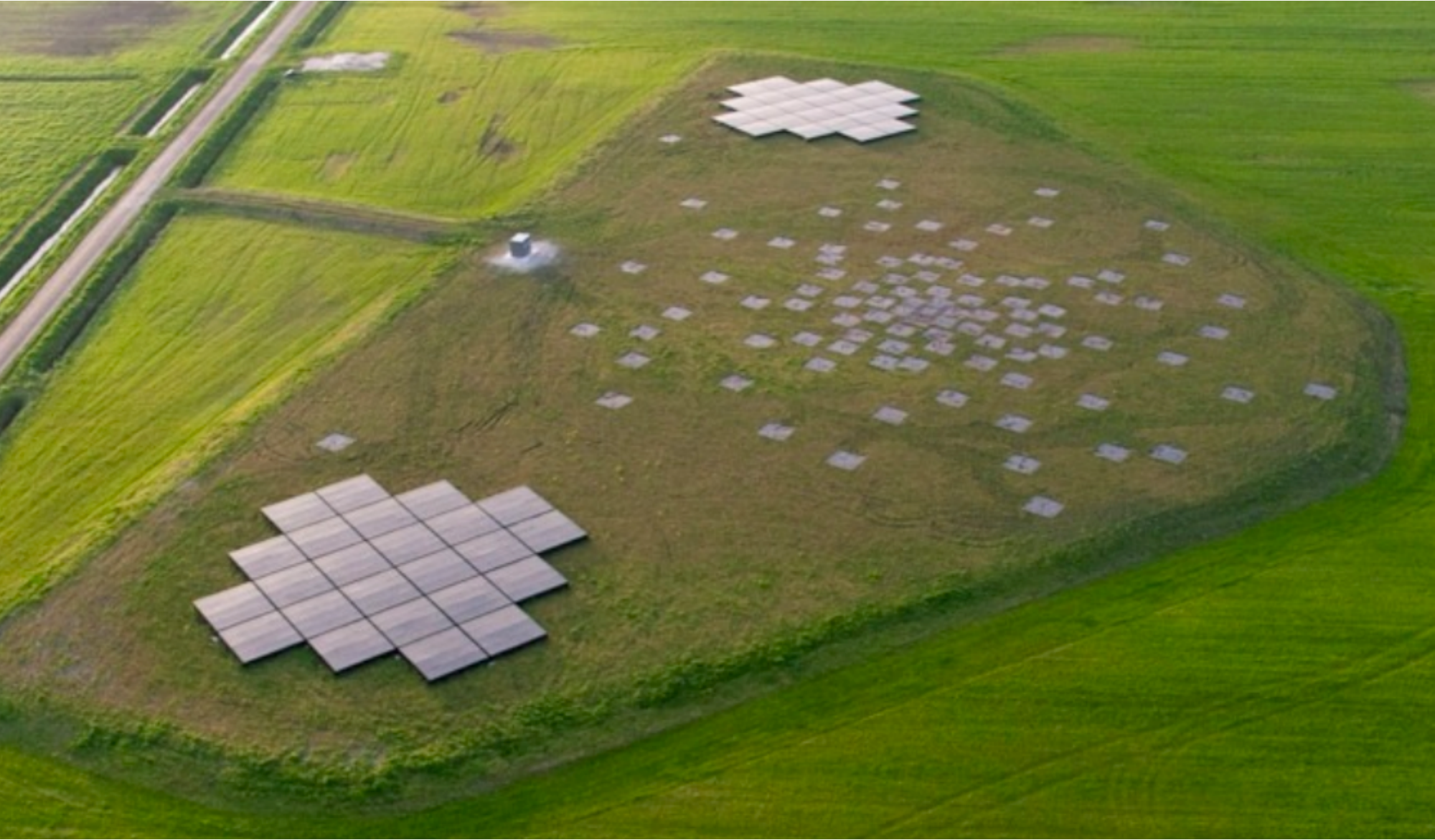}
\caption{An example of a LOFAR core station, showing the 96 simple dipoles that are combined as an LBA phased array, the 48 tiles that are combined as an HBA aperture array (split into $2\times24$~tile sub-fields) and the electronics cabinet in the centre. Note that for the remote and the international stations, the HBA field has all the tiles together.}
\label{figure:core}
\end{center}
\end{figure}

LOFAR is a new pan-European radio telescope that will observe the Universe between 10 and 250 MHz  to an unprecedented sensitivity and angular resolution. The array will consist of 22 core stations within a central 5 km near the village of Exloo (NL), 18 remote stations out to $\sim100$~km within the Netherlands and 8 international stations in Germany (5), France, Sweden and the United Kingdom. With baselines up to 1300 km and independent clocks at each station, LOFAR is truly a VLBI instrument. An example of a LOFAR core-station is shown in Figure \ref{figure:core}. Each of the LOFAR stations has a Low Band Antenna (LBA) and a High Band Antenna (HBA) field. The LBA fields are made up of 96 simple dipoles that are combined as a phased array and are sensitive between 10 and 80 MHz (i.e. before the onset of the FM band). The HBA fields are made up of either 48 or 96 tiles that form an aperture array system, which is sensitive between 110 and 250 MHz. For the core-stations, the HBA field is split into two sub-fields, whereas for the remote and international stations all of the HBA tiles are grouped together. Unlike the typical parabolic dish radio telescopes that are used as part of the EVN, the LOFAR stations have no moving parts. Observations towards a particular target are made by coherently adding the signals of the incoming radiation in the direction of interest. This beam-forming is carried out by computing and is highly flexible. For example, the LOFAR system has 244 sub-bands that can be used as a single beam with a total bandwidth of 48 MHz. Alternatively, the sub-bands can be used to form up to 244 different beams on the sky simultaneously, but with a smaller bandwidth per beam. The data from the individual stations are connected via optical fibres to the IBM BlueGene/P supercomputer in Groningen, where the correlated visibilities are produced. The huge fractional bandwidth, the large collecting area and the possibility of multi-beam observations makes LOFAR an ideal instrument for carrying out surveys of the low frequency sky. The main science goals of LOFAR are to carry out deep extragalactic surveys, probe the epoch of reionization, study cosmic magnetism, investigate the transient sky, study solar physics and observe ultra high energy cosmic rays. LOFAR will also provide an important testbed for the technologies (hardware and software) that will be used for the Square Kilometre Array.

The commissioning of the LOFAR imaging system has been ongoing since the first 3 working stations became available in the summer of 2009. However, given the environment where LOFAR is built and the complexity of the data analysis, new software programs have also needed to be developed over this period to reduce the data. The first step of the process is the removal of radio frequency interference (RFI) and the data compression (time and frequency). The RFI is mitigated using several algorithms (e.g. Offringa et al. \cite{offringa}), and we find at most $10\%$ of the data are flagged over a full dataset, with about 1--2\% of the data flagged from a typical subband. The RFI can consist of narrow and broadband interference in both frequency and time. The calibration of LOFAR data is also complex. The large field of view and the effect of the ionosphere means that a single calibration solution can no longer be used, and multiple direction dependent gains must be calculated. This is particularly important for the very bright off-axis sources that can interfere with the target. The system gain also varies for a given source due to the change in the source projection relative to the station. The large fields of view (7 to 1700 deg$^{2}$) over the full frequency range will also require advanced imaging techniques to be employed, either through image faceting or a large number of w-projections, or a combination of both. LOFAR will also generate staggeringly large data volumes, for example, a 6 hour dataset with 25 stations using 1 second visibility averaging will be 15~TB in size. Therefore, to analyze the data, an imaging pipeline is being developed that will take the raw data from the IBM BlueGene/P correlator, remove the RFI and compress the data, calibrate the visibilities, make images and generate source catalogues.

As an initial test of the LOFAR imaging capabilities with a partially completed array, we have been carrying out observations of a number of interesting objects, ranging from a large cluster halo, an active galaxy and an extended radio galaxy to wide-field images of the region around a dominant point-source. In these proceedings, we only present the data and an image that have been made for Cygnus A. For other examples of imaging with LOFAR, please see Heald et al. \cite{heald}.

\section{Imaging of Cygnus A}

The radio galaxy Cygnus A (3C~405; $z=0.056$) is one of the brightest sources in the LOFAR sky. It is one of the objects that makes up the `A-team'; a small number of very strong radio sources that can potentially contaminate every LOFAR observation. To reduce the effect of these sources on our science observations, we must image each member of the A-team over the LOFAR observing band and remove them from the {\it uv}-data of our target fields. The large luminosity and small distance to Cygnus A also makes this source interesting for studying the properties of AGN, for example, to investigate feedback processes. Its complex double structure also makes for an excellent commissioning target to test the LOFAR system.

\begin{figure}
\begin{center}
\includegraphics[angle=0,width=9.5cm]{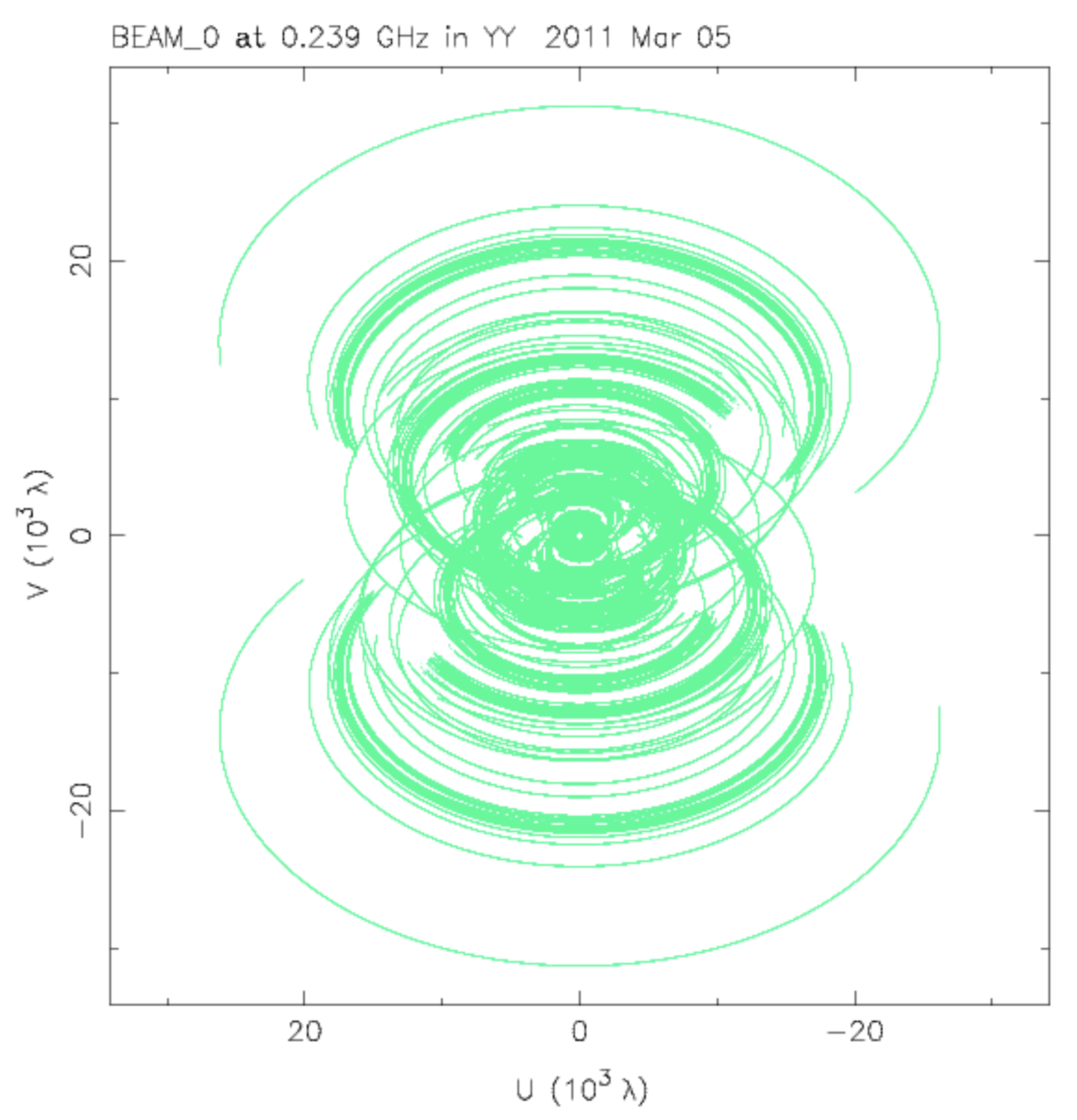}
\caption{The {\it uv}-coverage of a single sub-band observation of Cygnus A at 239 MHz with LOFAR. The integration time for this observation was 15 hours.}
\label{figure:uvplot}
\end{center}
\end{figure}

\begin{figure}
\begin{center}
\includegraphics[angle=0,width=13cm]{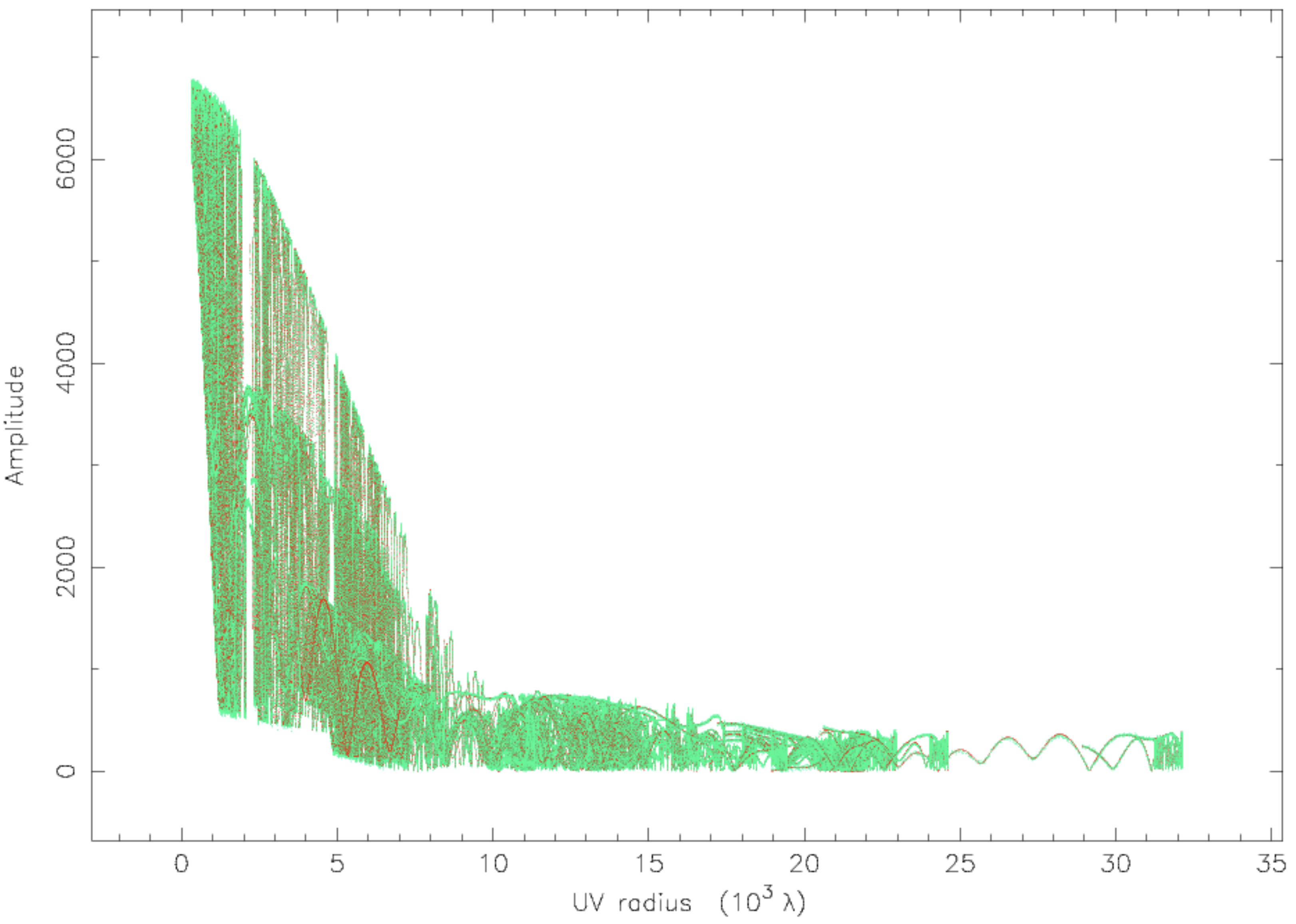}
\caption{The amplitude of the visibilities (green) as a function of uv-distance for Cygnus A. The model (red) is composed of $\sim6700$ clean components. Further developments will allow the model to be represented by shaplets and will include information about the frequency dependent structure.}
\label{figure:vis}
\end{center}
\end{figure}

\begin{figure}
\begin{center}
\includegraphics[width=\textwidth]{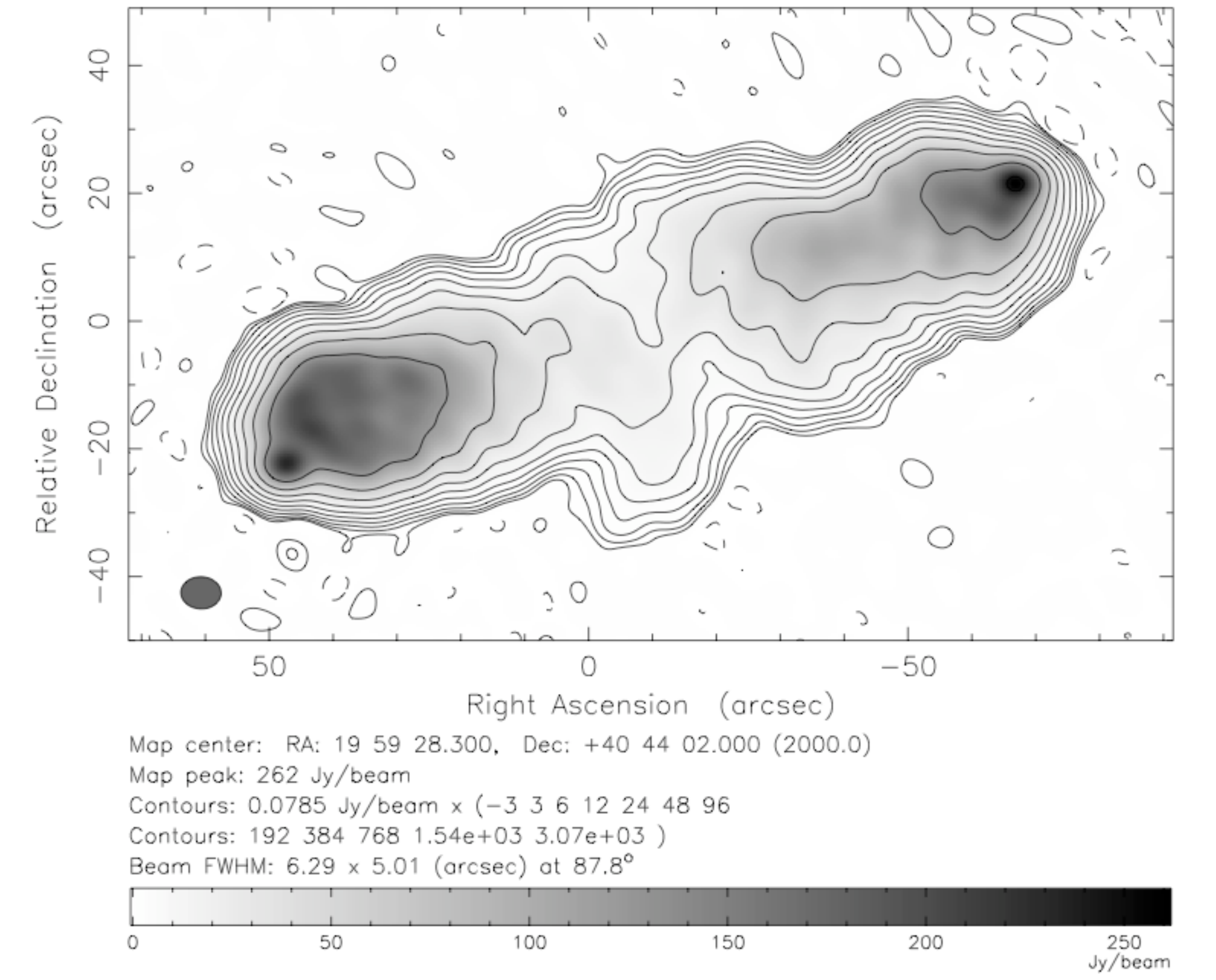}
\caption{The uniformly-weighted LOFAR image of Cygnus A at 239 MHz. The expected double-lobe structure, with the bright hot spots at the ends of the lobe and counter-lobe are observed. The dynamic range of this image is $\sim$3340.}
\label{figure:cyga}
\end{center}
\end{figure}

Cygnus A was observed for 15 hours in the LOFAR high band at $\sim$240 MHz on 2011 March 5. The array consisted of 19 core stations and 7 remote stations. The international stations were excluded from the array because the necessary VLBI techniques that are needed to analyze these kinds of data are still being developed - see the contribution by Wucknitz to these proceedings for some excellent examples of VLBI with LOFAR. The resulting {\it uv}-coverage of a single sub-band (0.2 MHz bandwidth) at 238.684~MHz is shown in Figure \ref{figure:uvplot}. The imaging, self-calibration and data editing could be carried out using {\sc difmap} because Cygnus A dominates the {\it uv}-data and because Cygnus A was in the centre of the beam throughout the observation; for other targets more complex calibration that includes direction dependent gain solutions and proper beam modelling is required. In Figure \ref{figure:vis}, we show the calibrated visibility data for Cygnus A. The zero-spacing flux-density was estimated to be 6740~Jy, which was calculated using the known spectral index of Cygnus A between 74 and 327 MHz, as measured with the VLA. The data quality is excellent for this strong source. The uniformly weighted image of Cygnus A is shown in Figure \ref{figure:cyga}. The image shows the expected double lobe structure that has been observed from Cygnus A at other frequencies (for example, note the similarity to the 327 MHz image shown by Lazio et al. \cite{lazio_et_al_2006}). We also see complex structure within the lobes and in the space between them, for example, the steep-spectrum hot spots in the lobe and counter-lobe are clearly detected. The rms noise in the image is 78.5 mJy~beam$^{-1}$, giving a dynamic range of $\sim$3340 (c.f. with the dynamic range of $\sim$4700 that is achieved with the VLA at 327 MHz). The limiting factor in the dynamic range is the cleaning. This image is an excellent example of what can be achieved with an aperture array system. The model of Cygnus A (see Figure \ref{figure:vis}) comprises $\sim$6700 clean components. In the future, the model will instead be parameterized using shaplets in order to speed-up the calibration process, and will also contain information about the frequency dependent structure of the source.

\section{Future outlook}

The data presented here are based on only a partially completed LOFAR. As of March 2011, there are 20 core stations, 7 remote stations and 5 international stations that are validated and are being used regularly as part of the array. The final core stations will be completed before Summer 2011, and the remaining remote stations are expected to come online sometime soon after then. There has also recently been improvements to the individual station calibration tables and the LOFAR beam server, which have seen an increase in the sensitivity and stability of the system. Also, the complex modelling of the LOFAR beam has been implemented and is now being tested in our calibration routines. The current challenges that are being faced by the Imaging Working Group are the removal of the bright A-team sources, like Cygnus A, and the automated processing of the large data volumes with an imaging pipeline. Once these problems have been overcome, then we will be in a position to start producing science quality data.

\acknowledgments{The work described here relies on the efforts of many people who have been involved in the LOFAR project. We gratefully acknowledge the huge amount of work done by the entire LOFAR hardware and software development teams. We also thank the much larger group of commissioners who take part in the LOFAR Imaging Busy Weeks, and have been instrumental in making progress in successfully calibrating LOFAR data. These efforts have also led to the development of the LOFAR Imaging Cookbook. Chiara Ferrari acknowledges financial support by the Agence Nationale de la Recherche through grant ANR-09-JCJC-0001-01.

LOFAR, the Low Frequency Array designed and constructed by ASTRON, has facilities in several countries, that are owned by various parties (each with their own funding sources), and that are collectively operated by the International LOFAR Telescope (ILT) foundation under a joint scientific policy.}


\begin{thebibliography}{99}
	\bibitem{offringa} A. R. Offringa et al., \emph{Post-correlation radio frequency interference classification methods}, \emph{MNRAS} {\bf 405} (2010) 155.
	\bibitem{heald} G. H. Heald et al., \emph{Recent LOFAR imaging pipeline results}, \emph{PoS} (2010) 57.
	\bibitem{lazio_et_al_2006} T. J. W.~Lazio et al., \emph{Cygnus A: A long-wavelength resolution of the hot spots}, \emph{ApJ} {\bf 624} (2006) L33.

\end{thebibliography}
\end{document}